\documentstyle[11pt]{article}
\begin{document}

\begin{center}{
{\Large \bf  Intercalation and Staging Behavior in Super-Oxygenated
$\rm La_2CuO_{4 + \delta}$}\\
\mbox{}\\

B.\@O.\@ Wells$^1$, R.\@J.\@ Birgeneau$^1$, F.\@C.\@ Chou$^1$,
Y.\@ Endoh$^2$, D.\@C.\@ Johnston$^3$,
M.\@A.\@ Kastner$^1$,  Y.\@S.\@ Lee$^1$,\\ G.\@ Shirane$^4$, J.\@M.\@
Tranquada$^4$, and K. Yamada$^2$ }\\
\end{center}
\mbox{}\\

\noindent $^1$ Department of Physics and Center for Materials Science and
Engineering, Massachusetts Institute \mbox{ \ \ } of Technology, Cambridge, MA,
02139, USA\\
\noindent $^2$ Department of Physics, Tohoku University, Aramaki Aoba, Sendai,
980-77, Japan\\
\noindent $^3$ Ames Laboratory and Department of Physics and Astronomy,
Iowa State University, Ames, IA 50011, USA\\
\noindent $^4$ Department of Physics, Brookhaven National Laboratory, Upton, NY
11973, USA \\

\begin{abstract} A high temperature electrochemical oxidation process
has been used to produce large single crystals of $\rm La_2CuO_{4
+ \delta}$ suitable for neutron scattering experiments.  Below room
temperature the oxygen-rich phases have structural superlattice
scattering peaks which indicate new periodicities ranging from 2 to
6.6 layers perpendicular to the copper oxide planes.  A model
structure originally proposed for $\rm La_2NiO_{4 + \delta}$ can
account for the superlattice peaks as a result of anti-phase domain
boundaries between different tilt directions of the CuO$_6$ octahedra.
Within this model, the changes in CuO$_6$ tilt directions are induced
by segregated layers of interstitial oxygen which order in a manner
similar to intercalants in graphite.  This structural model thus
clarifies previous work and establishes $\rm La_2CuO_{4 + \delta}$ as
a unique lamellar superconducting system with annealed disorder.\\
\end{abstract}

\noindent PACS Index Nos: 74.72.Dn, 74.25.Dw, 74.80.Dm, 61.12.-q
\newpage

\section{Introduction}

\hspace{.20in} One of the central questions concerning the physics of high
temperature superconductors is how the physical properties evolve as a
function of doping.  The planar copper oxides are antiferromagnetic
insulators when undoped.  The addition of charge carriers makes them
high temperature superconductors.  For example $\rm La_{2-x}Sr_xCuO_4$
is an insulating antiferromagnet for x=0 and a high temperature
superconductor with a maximum T$_c$ of about 38 K for x=0.15.  An
alternative method for doping $\rm La_{2-x}Sr_xCuO_4$ is to add extra
oxygen to form $\rm La_2CuO_{4 + \delta}$.  Very high pressures are
needed to oxidize $\rm La_2CuO_4$ by annealing in oxygen at high
temperatures, with pressures as high as 25 kbar needed to obtain
single phase superconducting samples.\cite{Zhou1} However, by using
electrochemical techniques it is possible to prepare heavily
oxygenated samples with $\delta$ as large as
0.12.\cite{Wattiaux,Chou1} $\rm La_2CuO_{4 + \delta}$ is
superconducting for large values of $\delta$ with a maximum of T$_c$
of 45 K.\cite{Johnston}

\hspace{.20in} Much of the interest in $\rm La_2CuO_{4 + \delta}$ is focused on
the observation of an intrinsic phase separation of the oxygen
dopants.  For oxygen contents such that 0.01 $\leq \delta \leq 0.06$
and temperatures below 300 K, $\rm La_2CuO_{4 + \delta}$ spontaneously
separates into metallic, oxygen-rich and insulating, oxygen-poor
regions.\cite{Jorgensen,Radaelli1} This phase separation is not a well
understood phenomenon.  In particular, the structure of the
oxygen-rich phase is not known.  Undoped $\rm La_2CuO_4$ has a fairly
simple, body centered tetragonal structure at high temperatures.  At
530 K there is a tetragonal to orthorhombic transition which is
characterized by a tilting of the octahedra made up of the Cu atom and
its six nearest neighbor oxygens around the tetragonal (1,1,0)
axis.\cite{Grande,Birg1} Most previous studies \cite{Radaelli2,Vaknin}
have reported that at low temperatures the oxygen-rich phase has, on
average, a simple face centered orthorhombic structure, with no
tilting of the CuO$_6$ octahedra.  The occurance of such a structure
is surprising because it is not clear how the interstitial oxygen
would suppress the tilting of the octahedra, or why the untilted
structure would be orthorhombic. There have been reports of extra
superlattice reflections in both neutron scattering \cite{Radaelli2}
and electron diffraction \cite{Tak} but these have not been understood
in terms of the structure of $\rm La_2CuO_{4 + \delta}$.

\hspace{.20in} In this paper we report a neutron scattering investigation of
single crystals of $\rm La_2CuO_{4 + \delta}$. We have found that the
low temperature structure of $\rm La_2CuO_{4 + \delta}$ is
orthorhombic with a superlattice structure along the c axis. The
diffraction is consistent with a structural model previously proposed
for $\rm La_2NiO_{4 + \delta}$.\cite{Tran} In this model the CuO$_6$
octahedra are tilted. Within the planes, the tilts are ordered in the
same manner as in undoped $\rm La_2CuO_4$ and perpendicular to the
planes the tilts have the same local ordering. The structure of $\rm
La_2CuO_{4 + \delta}$ with $\delta \geq$ 0.06 differs from that of
$\rm La_2CuO_4$ in that the doped material contains broad antiphase
domain boundaries regularly spaced along the c axis, perpendicular to
the CuO$_2$ planes. The antiphase domain boundaries consist of layers
across which the direction of the CuO$_6$ tilts are reversed. Within
this model the antiphase boundaries are presumed to be caused by
ordered layers of interstitial oxygen.  This one-dimensional ordering
of the extra oxygen is similar to the staging behavior of halogens or
alkalis intercalated into graphite.\cite{Safran} Following the
literature on intercalated graphite, we use the term staging to
describe the c axis modulation of the undoped structure; stage n
refers to an induced periodicity of n CuO$_2$ host layers. $\rm
La_2CuO_{4 + \delta}$ thus presents an interesting contrast with $\rm
La_{2-x}Sr_xCuO_4$ since the dopants which contribute the charge
carriers give rise to annealed rather than quenched disorder. This
makes $\rm La_2CuO_{4 + \delta}$ a particularly rich system for
studies of transport, magnetic, and superconducting properties.

\hspace{.20in} This paper is organized into several sections as
outlined below.  Section 2 describes the sample preparation and
measurement techniques.  Section 3 presents our neutron measurements
that determine the structure of the electrochemically oxygenated
samples. In section 4 we summarize the model of Tranquada {\it et al.}
\cite{Tran} and show that it is consistent with the neutron data.
Section 5 compares our results with previous experiments on $\rm
La_2CuO_{4 + \delta}$. This compound is compared with $\rm La_2NiO_{4
+ \delta}$ in section 6.  In section 7 we present results on a crystal
which clearly shows both staging and phase separation. Section 8
summarizes the primary results and conclusions of this work.

\section{Experimental Procedures}

\hspace{.20in} Several single crystal samples with different oxygen
contents were used in this experiment. Five different crystals were
measured, one of which was studied at two different stages of
oxidation. The three crystals, A, B, and D, were grown at MIT by the
top seeded solution method.\cite{Hidaka} Sample C was grown at AT\&T
Bell Laboratories by the flux method.\cite{Cheong} Sample E was grown
at MIT using the travelling-solvent-floating-zone (TSFZ) technique.
This latter method generally produces purer crystals since it does not
involve the use of a crucible. All five crystals were
electrochemically oxygenated. The electrolysis for sample C took place
in a 1 N solution of NaOH in water. The electrolysis current was fixed
to values which restricted the voltage to less than 0.6 Volts with
respect to a Ag/AgCl reference electrode in order to avoid the
electrolysis of water. The oxidation took place at room temperature
and continued for two months. Sample C was designated as ``crystal B''
in reference \cite{Chou2} and further details of the preparation and
characterization of sample C are available there.\cite{Chou2} The
electrolysis of samples A, B, D, and E was carried out in the same
electrolyte but with a fixed voltage of 0.4V or 0.45V with respect to
the Ag/AgCl reference electrode. In addition, the electrolysis was
performed at elevated temperatures to speed the oxidation process.
Samples A1 and A2 refer to the same crystal A after different amounts
of electrolysis. Table 1 summarizes the material and electrolysis
parameters of the six samples.

\hspace{.20in}  The samples have been characterized with bulk magnetization
measurements in a Quantum Design SQUID magnetometer.  Figure 1 shows
magnetization measurements of sample B in a field of 0.5 T taken at
weekly intervals during the oxidation process.  The magnetization for
the as-grown crystal has a peak at 225 K indicating the N\'{e}el
temperature, T$_N$.  The peak in the susceptibility is actually caused
by the weak ferromagnetic transition which accompanies the
antiferromagnetism.\cite{Thio} At this point the sample shows no sign
of superconductivity above 5 K.  After one week, the peak in the
magnetization is much broader and is shifted to a lower temperature.
The center of this broad peak is near 150 K.  In addition, low
temperature, low field magnetization, not shown, shows a shielding
signal with an onset near 29 K indicative of superconductivity, which
is broad in temperature.  This suggests that after a week of
electrolysis, the crystal contains a distribution of oxygen contents
and does not show two phase coexistence, presumably because of kinetic
limitations.  After another week of oxidation, the large N\'{e}el peak
disappears and only a small residual N\'{e}el peak at 220 K remains.
At this point the shielding measurement gives a large signal for
superconductivity with a fairly broad transition and T$_c$=29 K at
onset.  (We note parenthetically here that superconducting T$_c$'s for
crucible-grown single crystals are typically somewhat lower than those
of nominally equivalent ceramics.)  The crystal now exhibits the
coexistence of two phases, an oxygen-poor phase with T$_N$=220 K and
an oxygen-rich superconducting phase.  The peak at T$_N$ continues to
shrink as the crystal is further oxidized.  Most of the change in the
macroscopic magnetization takes place in the first two weeks.  After
the completion of the electrolysis procedure, the remnant peak is
approximately 1/15 of the original N\'{e}el peak, indicating the
fraction of the sample that remains in the oxygen-poor phase.

\hspace{.20in}  Diffraction measurements were performed at beam lines H7 and H8
at the High Flux Beam Reactor at Brookhaven National Laboratory.  The
monochromators and analyzers were pyrolytic graphite (PG) set for the
(002) reflection.  Many different configurations were used in these
experiments.  Two PG filters were used to suppress reflections from
$\lambda/2$ and $\lambda/3$ neutrons.  Collimations used varied from
$40^{\prime}-40^{\prime}-S-80^{\prime}-80^{\prime}$ to measure weak
superlattice peaks in small samples to
$10^{\prime}-10^{\prime}-S-10^{\prime}-10^{\prime}$ for high
resolution measurements of fundamental peaks.  All of the samples
studied were wrapped with Al, sealed in an Al can with He exchange
gas, and cooled in a Displex closed cycle He cryostat.  Temperature
was measured with a Pt resistance thermometer and a Si diode attached
to the cold finger.

\section{Structural Data}

\hspace{.20in} Figure 2 shows a real space model of $\rm La_2CuO_{4
+ \delta}$.  Part a is the orthorhombic Bmab structure of the undoped
material and part b is the proposed modification of the structure for
oxygenated $\rm La_2CuO_{4 + \delta}$, in analogy to the corresponding
structure
\cite{Tran} in $\rm La_2NiO_{4 + \delta}$.  This will be discussed in
more detail below.  A map of the attendant reciprocal space indexed to
the orthorhombic Bmab unit cell is shown in Figure 2c.  In this paper
we use the notation {\bf a}, {\bf b}, {\bf c} to refer to the
orthorhombic Bmab real space axes, and H, K, L for reciprocal space
axes as well as general peaks at (H,K,L), with reciprocal lattice
constants {\bf a}$^{\ast}$, {\bf b}$^{\ast}$, and {\bf c}$^{\ast}$.
Neutron scattering measurements using a triple axis spectrometer are
restricted to a plane in reciprocal space.  Due to the {\bf a}-{\bf b}
twinning present in $\rm La_2CuO_4$, our scattering plane is a
superposition of the (H,0,L) and (0,K,L) planes.  The fundamental or
Fmmm peaks in this plane occur where both H and L are even or K and L
are even.  In addition, the Bmab phase has superlattice reflections
which occur for H=0, K odd, and L even.

\hspace{.20in} In Figure 3 we show the scattering profiles for samples C, A2, D
and E along the L axis in the vicinity of the (0,1,4) position of
reciprocal space.  Each scan shows several peaks.  In samples E, D,
and A2, the (0,1,4) Bmab peak is visible.  In addition, there are
peaks at non-integer values of L.  These peaks appear in pairs, in
which each member is split from the (0,1,4) position by an amount
$\Delta$.  Sample E has $\Delta = 0.15$, sample D has $\Delta = 0.18$,
sample A2 has $\Delta = 0.25$, and sample C has $\Delta = 0.25$, 0.33,
and 0.5.  Figure 2c indicates the positions of these peaks in
reciprocal space for $\Delta = 0.25$.  With $\Delta = 1/n$, the
superstructure peaks at larger $\Delta$ are at integer values of n
whereas those at smaller $\Delta$ are incommensurate; for sample E,
n=6.6, for sample D, n=5.5, for sample A2, n=4, and for sample C, n=2,
3 and 4.  Sample A1 also has n=5.5 and sample B has n=4.2.  The four
samples in Figure 3 illustrate the range of values we have detected
for $\Delta$, or n.

\hspace{.20in} Figure 4 shows the diffraction profile along L for sample B, a
sample with close to the stage 4 structure, in the vicinity of the
(0,3,2) Bmab superlattice peak.  The data in this figure cover a wider
range in L than shown in Figure 3.  Well-developed peaks are seen near
(0,3,2 $\pm \frac{1}{4}$), but any peaks at the positions of the
higher harmonics of (0,3,2 $\pm \frac{3}{4}$) are hardly detectable.
There does appear to be a very small peak at (0,3,1$\frac{1}{4}$)
which is about 1/100 of the intensity of the (0,3,1$\frac{3}{4}$)
peak.  This is the only higher harmonic peak that we have found
throughout the (H,0,L)/(0,K,L) zone.  We also have conducted an
extensive search for other superlattice peaks.  In particular, we have
searched for peaks that would correspond to a new in-plane ordering
including scans in H or K with L an integer or at a staging position
(L = integer $\pm$ 1/n) as well as scans in L at various values of H
or K.  We have found no other superlattice peaks within the
(H,0,L)/(0,K,L) zones.

\hspace{.20in} A practical way to distinguish scattering peaks from the
oxygen-rich and oxygen-poor phases is through the differing length of
the {\bf c} axis lattice constant.  A high resolution scan of the
fundamental (0,0,8) position shows two peaks indicating that there are
two phases with different lattice parameters present in the sample.
The values for the c lattice constants are shown in Figure 5a.  One
peak corresponds to the position of the (0,1,4) Bmab superlattice peak
while the other corresponds to the average of the (0,1,4 $\pm \Delta$)
peaks.  Thus the peaks at (0,1,4) and (0,1,4 $\pm
\Delta$) are from separate phases.  From the intensity of the two (0,0,8)
peaks, we can estimate the fraction of the sample in the Bmab and staged
phases.  For example, in sample B the ratio of (0,0,8) peak intensities
indicates that about 5\% of the sample is oxygen-poor.
This compares well with  the
relative strength of the N\'{e}el peak in the magnetization curve of Figure 1
which indicates that 7\% of the sample is oxygen-poor.  For the samples in
Figure 3, we find that sample E is 80\% oxygen-poor, sample D is 27\%
oxygen-poor, and sample A2 is 2\% oxygen-poor.  The Bmab peaks in samples D
and A2 look large in comparison to the staging peaks because they are much
narrower.  A high resolution scan of the (0,0,8) peak in sample C indicates
that several phases are present, but even the highest spectrometer resolution
is inadequate to distinguish them.

\hspace{.20in} Figure 5 shows the lattice constants we measure for samples A1,
B and C.  For samples A1 and B lattice parameters for both the
minority oxygen-poor and the majority oxygen-rich phase are given.
For sample C, we are not able to resolve peaks from the different
phases, so the given values are an average of the three staged phases
that are present.  There is a large expansion of the {\bf c} axis for
the staged samples compared to the undoped phases.  We expect such an
increase in volume to accommodate the doped interstitial oxygens.
However, there is only a slight increase in the c axis between the
stage 5.5 phase and the stage 4 phase and no further expansion for the
stage 2-4 sample.  Comparing the {\bf a} and {\bf b} axes for the
different phases shows that the introduction of extra oxygen causes a
contraction of the in-plane lattice constants, although, once again,
the differences in the in-plane lattice constants between the
oxygen-poor phase and the oxygen-rich phase is much greater than the
differences between differently staged phases.  The behavior of the
lattice constants in $\rm La_{2-x}Sr_xCuO_4$ is very similar.  For Sr
contents less than x $\sim$ 0.01, the {\bf a}-axis shrinks and the
{\bf c}-axis grows with increasing x, while for Sr contents greater
than x $\sim$ 0.01, that is, Sr contents corresponding to
superconducting phases, the $\frac{{\bf a} + {\bf b}}{2}$ and ${\bf
c}$ lattice constants are nearly independent of x.\cite{Birg2} As
discussed below, the lowest density oxygen-rich phase we observe
corresponds to $\delta \sim 0.06$ or a hole concentration equivalent
to $\rm La_{2-x}Sr_xCuO_4$ with x as high as 0.12. The exact hole
concentration ($p$) as a function of $\delta$ in $\rm
La_2CuO_{4+\delta}$ samples depends on the ionization state of the
intercalated oxygens and this is a matter of debate in the literature
with reported values of $p/\delta$ ranging from 1 to
2.\cite{Johnston,Radaelli2,Zhou2,Quijada}

\hspace{.20in}  The orthorhombic phase is stable to higher temperatures for
the phases with lower staging numbers.  Undoped $\rm La_2CuO_4$ has a
transition from a high temperature tetragonal phase (HTT) to a low
temperature orthorhombic phase (LTO) at approximately 530 K.  While we
did not raise the temperature high enough to observe the transition in
the oxygen-rich phases, the stage 5.5 sample has the smallest
orthorhombic strain at room temperature and should become tetragonal
at the lowest temperature.  Extrapolating from the data in Figure 5
indicates that the HTT to LTO transition is near 400 K in the stage
5.5 sample.  As the stage number decreases, the orthorhombic strain
increases, and presumably, the phase transition to the tetragonal
phase moves to higher temperatures.  Thus, initially the tetragonal to
orthorhombic transition temperature decreases with the introduction of
extra oxygen, but then rises as the staging number is lowered.  This
same behavior of the HTT to LTO transition has also been reported by
Radaelli {\it et al.}. \cite{Radaelli1} In $\rm La_{2-x}Sr_xCuO_4$ on
the other hand, the temperature of the tetragonal to orthorhombic
phase transition decreases monotonically with increasing Sr content,
and the material remain tetragonal at the lowest temperatures measured
for a Sr content of 0.2 or more.\cite{Birg1}

\hspace{.20in} The temperature dependence of the staging peaks indicates that
the various stagings are equilibrium phases.  Three different samples
include stage 4 phases: \ samples A2 and B are greater than 90\% stage
4 while sample C has a minority stage 4 phase.  Figure 6 shows the
temperature behavior of the (0,3,2 $\pm \frac{1}{4}$) peak for the
stage 4 phases in samples B and C.  The integrated intensity and the
in- and out-of-plane half-widths-at-half-maxima (HWHM) are shown.
These data are derived from fits of the in- and out-of-plane
directions of the stage 4 peak in sample B.  In the simplest model the
widths of these peaks are proportional to the inverse of the size of
the coherent patches of stage 4 phase.  Sample B has ordered domains
that are 120\AA \ wide in-plane and 60\AA \ out-of-plane while the
domains in sample C are 230\AA \ in-plane and 130\AA \ out-of-plane.
Here we have taken the domain size as the inverse of the de-convolved
HWHM.  It should be noted that the out-of-plane widths may also arise
in part from staging disorder, with the stage 4 peak representative of
a mixture of stages 3, 4 and 5.  Staging disorder is likely to be a
major contributor to the peak broadening for the stage 5.5 and 6.6
phases which more clearly represent mixtures of pure stages.  The
difference in domain size may reflect the quality of the initial $\rm
La_2CuO_4$ crystal, or alternatively, may indicate the equilibrium
domain size. The integrated intensity in the two crystals falls off in
the same manner with increasing temperature indicating an ordering
temperature of about 300 K.  As noted before, the stage 5.5 peak has a
similar temperature dependence to that of the stage 4 peak, ordering
near 300 K.  The stage 2 and 3 peaks remain ordered to higher
temperatures.  We have observed very little dependence of the widths
or intensities of these staging peaks on the rate at which the sample
is cooled, indicating that the time scale for ordering the staging
domains is smaller than the time over which we can cool the sample.
The cooling rate for the displex cryostats used in the neutron
scattering experiments is fairly slow, taking about 3 hours to cool to
10 K from room temperature.  As we point out below, in susceptibility
measurements, for which we can quench more quickly, the crystals show
indications of an ordering time of at least several seconds.

\section{The Structural Model}

\hspace{.20in}  In order to explain the diffraction peaks we use a
structural model previously presented for $\rm La_2NiO_{4 + \delta}$
by Tranquada {\it et al.}.\cite{Tran} To explain the model, we use
several simplifications.  We refer to CuO$_6$ octahedra when, in fact,
even in the undoped material the CuO$_6$ complex does not form a
perfect octahedron.\cite{Grande,Birg1} Futhermore, the true
distortions involved in the tetragonal to orthorhombic transition are
more complicated than simple rotations of the CuO$_6$ complex.
However, the symmetries produced by the structural distortions are the
same as those that would be produced by a rigid rotation of the
CuO$_6$ unit, and this description provides an easily understood
description of the various structures involved in this study.  In
addition, we describe the model as having absolutely sharp domain
boundaries and completely segregated sheets of oxygen.  In fact, it is
likely that the domain walls and oxygen sheets are spread across a few
atomic layers.  In some cases, we have strong evidence for such
smearing of the boundaries, as we point out below.

\hspace{.20in} At high temperatures undoped $\rm La_2CuO_4$ has the body
centered, tetragonal $\rm K_2NiF_4$ structure, with space group Immm, the HTT
phase.\cite{Birg1}  This structure consists of
sheets of $\rm CuO_6$ octahedra sharing
in-plane corners.  La ions are nearly coplanar with the apical oxygen
ions.  At 530 K there is a phase transition to an orthorhombic structure
with space group Bmab, the LTO phase.  This orthorhombic structure is
achieved by tilting the octahedra along a direction at 45$^{\circ}$ with
respect to the Cu-O in-plane bonds, that is, around the orthorhombic [100]
direction.  Because each in-plane O ion is shared between two $\rm CuO_6$
octahedra, adjacent oxygen-sharing octahedra must tilt in opposite directions
as illustrated in Figure 2.  The apical oxygens are not shared between $\rm
CuO_6$ octahedra in different layers so only weaker electrostatic forces
determine the alignment of tilts from layer to layer.  In the Bmab
structure the alignment of adjacent planes is such that all octahedra
with the same displacement along b tilt in the same direction.  This is
illustrated in Figure 2a.

\hspace{.20in} As noted above, the reciprocal space cell that we are
using is referred to a unit cell for the orthorhombic, Bmab structure.
This orthorhombic unit cell has twice the in-plane area of the primitive
unit cell for the HTT structure.  The scattering for the tetragonal phase
in the zone accessible in this experiment gives Bragg peaks at (H even,
K=0, L even).  A simple orthorhombic distortion, in which the only change
is a stretching of the {\bf b} axis, leads to what is known as the Fmmm
structure.  This structure gives the same scattering as observed in the
tetragonal case except that H $\neq$ K, and because of twinning, both
zones (H,0,L) and (0,K,L) are observed in our scattering geometry.  The
Bmab superstructure, which is generated by the displacements attendant
with the octahedra tilts, gives rise to superlattice peaks in the neutron
scattering in the (H,0,L)/(0,K,L) zone at positions (H=0, K odd, L
even).  The oxygenated crystals show a splitting of the Bmab superlattice
peaks along L, but we have failed to observe  new peaks at
values of H$\neq$0.  Qualitatively,
the scattering pattern reveals the following about the structure of $\rm
La_2CuO_{4 + \delta}$:
\begin{enumerate}
\item a new periodicity along {\bf c}, n times the orginal {\bf c} lattice
constant,
\item the change involves the octahedra tilts, and
\item there are no displacements along the [1,0,0] direction that lead to
new symmetries.
\end{enumerate}
These points are included in the model structure for $\rm La_2CuO_{4 +
\delta}$ that we discuss below.

\hspace{.20in}  The underlying physics of this model is as follows.  It
has been suggested by Chaillout {\it et al.} \cite{Chaillout1} that
the interstitial oxygen atoms sit in the (1/4,1/4,1/4) sites of the
orthorhombic unit cell.  This site is betwen pairs of layers
containing La and apical O atoms, directly above an in-plane oxygen.
A single negatively charged oxygen ion in such a site repels nearest
neighbor apical oxygen ions and attracts La ions.  A local tilt flip
of the $\rm CuO_6$ octahedron either above or below the excess oxygen
ion creates a large interstitial position and maximizes the average
distance between the extra oxygen ion and surrounding oxygen ions.
Within a layer, the flipped tilt pattern propagates because of the
shared corner oxygens.  In turn, this creates an entire array of
favorable interstitial sites in the same plane as the original
interstitial oxygen.  We illustrate such an arrangement in Figure 2b.
Other interstitials fills this plane, but electrostatic repulsion
between oxygen ions or the number of available sites limits the
density within any one interstitial layer.  The spacing between
interstitial layers is governed by long range strain
energies.\cite{Safran} Between planes of interstitial oxygen ions, the
structure is still Bmab, but in addition, anti-phase boundaries
between the two possible Bmab orientations exist at every layer of
interstitial oxygen ions.

\hspace{.20in}  We have calculated the structure factors for the model
proposed above in order to compare the scattering it predicts with
experiment.  The calculations have been performed for a single,
expanded unit cell that represents the full symmetry of a perfectly
staged sample.  For example, for a stage 4 phase, our unit cell is the
same as for the orthorhombic Bmab phase in the {\bf a} and {\bf b}
directions, but it is 4 Bmab unit cells long in the {\bf c} direction.
This large unit cell contains 8 $\rm CuO_2$ layers, the first four
with octahedral tilts in one direction and the next four with the
opposite tilt direction.  The effect of this variation on the Bmab
structure is that the Bmab superlattice peaks at (0, K odd, L even)
have zero amplitude in the structure factor and are replaced by peaks
displaced along the L axis by $\pm 0.25 {\bf c}^{\ast}$ and $\pm 0.75
{\bf c}^{\ast}$.  The positions of the allowed peaks and the general
intensity variation of our calculated structure factors throughout
reciprocal space are in agreement with the data.  However, the
octahedra must be distorted to match quantitatively the intensities of
the peaks throughout the zone.  Determining this distortion
unambiguously is impossible with the few superlattice peaks we have
measured.  For our calculation, we have used the atomic positions
given by Zolliker {\it et al.} \cite{Zolliker} for the LTO, undoped
phase.  These correspond to $\rm CuO_2$ octahedral tilts of 5.5
degrees.

\hspace{.20in} The structure factor calculation for this model predicts peaks
at positions that correspond to higher odd harmonics of the staging
peaks.  That is, for the stage 4 sample, the calculation gives first
harmonic peaks at (0, K odd, L = even $\pm$ 1/n) and third harmonic
peaks at (0, K odd, L = even $\pm$ 3/n).  The third harmonic peaks are
predicted to be weak, but definitely stronger than any seen in the
data.  For example, the calculation predicts that the strongest third
harmonic peak, over the range of reciprocal space that we have
studied, is the (0,3,1$\frac{1}{4}$) peak which is 1/5 the calculated
intensity of the (0,3,1$\frac{3}{4}$) peak.  As shown in Figure 4
however, the (0,3,1$\frac{1}{4}$) peak is only about 1/100 the
intensity of the (0,3,1$\frac{3}{4}$) peak.  We can account for this
suppression of the higher harmonic peaks by allowing a variation in
the degree of tilt of the $\rm CuO_6$ octahedra.  The original model
allows only for a tilt of $+\Theta$ for the n layers on one side of
the antiphase domain boundary and $-\Theta$ on the other. Thus for a
given {\bf b} coordinate, the $\rm CuO_6$ tilt angle forms a square
wave along the {\bf c} direction.  If the $\rm CuO_6$ tilt angle forms
a sine wave instead, then the scattering at higher harmonics of the
staging periodicity is suppressed.  For a stage 4 structure such a
sine wave means that with each stack of four $\rm La_2CuO_4$ layers,
the two layers of $\rm CuO_6$ octahedra bordering the domain
boundaries have a smaller tilt angle than the octahedra in the two
layers not adjacent to the boundary.  The structure factor for this
configuration with tilts of 5.5$^{\circ}$ and 2.75$^{\circ}$ for the
inner and border $\rm CuO_6$ octahedra respectively, predicts that the
intensity of the (0,3,1$\frac{1}{4}$) peak is 1/100 of the intensity
of the (0,3,1$\frac{3}{4}$) peak, in agreement with the data.

\hspace{.20in} While the peaks displaced along L from the Bmab positions can be
explained merely by the changing tilt directions of the $\rm CuO_6$
octahedra, the intercalated layers of oxygen should themselves produce
scattering peaks at (0,0,2z/n), with n the staging number and z an
integer.  We do not observe such peaks above the background for
samples A and B.  In sample C, we do see peaks at (0,0,z) with z odd.
This would correspond to peaks from the excess oxygen in stage 2.
However, our structure factor calculation shows the observed peaks are
stronger than expected for scattering from the excess oxygen and may
correspond to other effects such as expansion of the layer in which
the excess oxygen is inserted.  There are no detectable peaks at the
positions for scattering from the intercalated oxygen for the stage 3,
4 or 5.5 phases.  Thus, we have not been able to detect a scattering
peak directly from the excess oxygen.  We have performed structure
factor calculations for our model structure in order to determine
whether the peak for scattering directly from the interstitial oxygen
ought to be strong enough to detect.  Assuming an oxygen content of
4.0625 for a stage 4 sample, corresponding to 1/4 of the large
interstitial sites occupied, and that all of the excess oxygen atoms
are confined to the antiphase domain boundary layer, the (0,0,2z/n)
peaks should be approximately 1/100 of the (0,3,1$\frac{3}{4}$) peak
shown in Figure 4.  Such peaks should be just barely observable; the
absence of these planar oxygen diffraction peaks argues for either
staging disorder or a broader distribution of the intercalated oxygen.

\section{Comparison to Other Results}

\hspace{.20in}  Several other groups have proposed models for the structure of
the oxygen-rich phases of $\rm La_2CuO_{4 + \delta}$.  The model we
have used is different from any used before for $\rm La_2CuO_{4 +
\delta}$, but much of the actual data presented in the previous papers
are consistent with both our data and the staging model.  Our neutron
scattering experiments provide a direct measure of the structure.  We
have used crystals with as great a range of oxygen contents as any
previously studied, and our large crystals allow a higher resolution
study than earlier experiments.  This has allowed us to identify the
weak, but essential, superlattice peaks which were missed or
misidentified in earlier work.

\hspace{.20in}  A detailed comparison  of our results with previous papers
indicates that much of the earlier data are consistent with the
oxygen-rich phases being staged.  In particular, those papers which
report that the oxygen-rich structure is Fmmm most likely simply
missed the weak, broad staging superlattice
peaks.\cite{Radaelli2,Vaknin} These peaks would be particularly
difficult to observe in powder samples.  We do find that the Bmab
superlattice peaks disappear in the oxygen-rich phase in agreement
with Radaelli {\it et al.}.\cite{Radaelli2} The staged peaks that
replace the Bmab diffraction peaks can vary in width and intensity
depending upon the overall oxygen content and the degree to which the
staged tilt pattern is able to order.  Vaknin {\it et al.}
\cite{Vaknin} report a broadening and weakening of the Bmab
superlattice peak at 260 K.  They themselves note that this may be due
to an unresolved splitting of the Bmab peak.  Radaelli {\it et al.}
\cite{Radaelli2} report neutron scattering measurements on a small
single crystal doped beyond the phase separation region.  The small
size of their crystal limits the resolution that they are able to use
and complicates the comparison to our data.  They report several
different sets of superlattice peaks.  At least one set appears to be
consistent with a stage 3 model.  We do not detect any sign of the
other superlattice peaks they report.

Above room temperature the staging peaks for the stage 4 and higher staged
phases disappear.  No other superlattice peaks appear at this temperature.
Thus, above room temperature the average structure is Fmmm.  In this case, the
tilts presumably still exist, but are disordered.  Indeed, this staging
order-disorder transition warrants further study.  If the tilt angle has a
sine-wave distribution,the transition occurs at the temperature at which the
amplitude of the sine-wave goes to zero.  It is possible that in some of the
previous experiments this Fmmm structure was locked in at lower temperatures
by quenching the sample; this could also occur if a large number of defects
pin the direction of the $\rm CuO_6$ tilts and do not allow them to order.  A
separate, early neutron scattering experiment by Chaillout {\it et al.}
\cite{Chaillout2}
claims that the oxygen-rich phase is a distinct structure with the same Bmab
symmetry as the oxygen-poor and undoped phase.  Neither this experiment nor
those listed above which claim that the oxygen-rich phase is Fmmm see
diffraction peaks that could be consistent with such a structure.

\hspace{.20in} Hammel {\it et al.} \cite{Hammel} use results from NMR and NQR
experiments to propose a model in which the magnitude of the CuO$_6$
octahedral tilts vary within a plane.  In this model there is a continuous
variation of the tilt angle of neighboring CuO$_6$ octahedra displaced along
the b axis from a maximum value of $ + \Theta_{max}$ to a minimum of $ -
\Theta_{max}$.  After reaching the value $ - \Theta_{max}$, the next
octahedron flips to $+ \Theta_{max}$ and a favorable interstitial site
is formed.  This arrangement of $\rm CuO_6$ tilts is not compatible
with our structural diffraction data.  We see no evidence for in-plane
scattering peaks that would arise if such an arrangement of the $\rm
CuO_6$ octahedral tilts were ordered.  However, our data do indicate
that there is some distribution of tilt angles.  As described above,
in the stage 4 sample with a sine wave distribution of tilts, two
different $\rm CuO_6$ species are present, a small tilt in layers
adjacent to the antiphase domain boundary and a larger tilt in those
layers further from the boundary.  Our neutron diffraction experiments
only determine the average structure.  If the values for the tilt
angles were, for example, $2\frac{3}{4}$ and $5\frac{1}{2}$ degrees,
with a variation of plus or minus one degree for either, then there
would be a fairly uniform distribution of local $\rm CuO_6$ octahedral
tilts, consistent with the NMR-NQR data.

\hspace{.20in} The optical properties of our sample C have been measured
by Quijada {\it et al.}.\cite{Quijada} They have found that the 500
cm$^{-1}$ mode of the apical oxygen atoms splits into two distinct
components in the oxygen rich phase. It is difficult to make a direct
comparison of this near surface measurement to our bulk structure
determination for such a multi-phase sample. However, we speculate
that the two components of the apical oxygen mode correspond to those
apical oxygen atoms adjacent to the intercalated oxygen layers and
those not neighboring the excess oxygen layers.

\section{Comparison to $\rm La_2NiO_{4 + \delta}$}

\hspace{.20in} $\rm La_2NiO_{4 + \delta}$ exhibits many different structural
phases as a function of the oxygen content as has been reported by
Tranquada {\it et al.}.\cite{Tran} For oxygen contents 0.058 $< \delta
<$0.125, $\rm La_2NiO_{4 +
\delta}$ has staged phases with scattering profiles very similar to those we
measure for $\rm La_2CuO_{4 + \delta}$.  Both $\rm La_2NiO_{4 +
\delta}$ and $\rm La_2CuO_{4 + \delta}$ are tetragonal at high
temperatures.  In $\rm La_2NiO_{4 + \delta}$ the staging peaks appear
at the HTT to LTO phase transition at T $\leq$ 290 K.  In $\rm
La_2CuO_{4 + \delta}$ for stage 4 and higher, the staging peaks order
near room temperature, but this is well below the tetragonal to
orthorhombic structural phase transition temperature.  We have not
determined whether the stage 2 and 3 phases order independently of the
HTT to LTO structural transition.

\hspace{.20in} Several different staged phases have been reported in various
samples of $\rm La_2NiO_{4 + \delta}$.  Slowly cooled, staged samples include
the following:
\begin{enumerate}
\item a single superlattice indicating nearly pure stage 2 for $\delta$=0.105,
\item two superlattices indicating nearly pure stage 2 and stage 3 phases for
$\delta$=0.085,
\item two superlattices indicating one phase with mixed stage 3 and 4 and a
lightly oxygenated phase with the low temperature tetragonal structure
(LTT) for $\delta$=0.06.
\end{enumerate}

In all of these samples, broad scattering peaks that correspond to intermediate
staging appear if the samples are cooled quickly.  We have not seen such a
dependence on cooling rate for $\rm La_2CuO_{4 + \delta}$.  The time scale for
ordering into staged structures appears to be less than one hour in $\rm
La_2CuO_{4 + \delta}$, but is several hours in $\rm
La_2NiO_{4 + \delta}$.\cite{Lorenzo} In addition, we
detect a wider range of stagings than has been seen in the
nickel oxide materials.  In $\rm La_2CuO_{4 + \delta}$ we have seen a range of
stages from 2 to 6.6 while in $\rm La_2NiO_{4 + \delta}$, only stagings from 2
to 3.5 have been detected.

\hspace{.20in} Most of the crystals of $\rm La_2CuO_{4 + \delta}$ and
$\rm La_2NiO_{4 + \delta}$ studied show coexistence of phases of
different pure integer stagings or a staged phase plus an oxygen-poor
phase.  Single superlattice peaks that represent mixed stage phases
occur mostly for the larger staging numbers observed in either
compound.  The free energy difference between phases with large
staging numbers must be very small so it may be particularly difficult
to reach an equilibrium state for such configurations.

\section{Phase Separation}

\hspace{.20in} Further experiments are necessary before we can reach strong
conclusions about the relationship between staging and the separation
of oxygen-rich and oxygen-poor phases.  In this section we point out
the implications of the work carried out so far.  In only one of the
samples studied to-date, sample E, have we been able to detect
definite evidence of an equilibrium, temperature driven phase
separation.  The magnetization at high field revealing the weak
ferromagnetic transition \cite{Thio} associated with the
antiferromagnetism of the undoped phase is shown in the top panel of
Figure 7.  The shielding signal for superconductivity in the
oxygen-rich phase is shown in the lower panel.  The temperature
dependence of the high field magnetization is very similar for the
pre- and post-electrolysis samples.  Assuming that the per site
magnitude of the weak ferromagnetic moment is similar as well, the
weak ferromagnetic signal strength at low temperatures provides a
measure of the volume of the sample that is oxygen-poor.  Using this
criterion, we find that below the phase separation temperature, sample
E is 20\% oxygen-rich and 80\% oxygen-poor.  We also see from the
shielding measurement that the superconducting transition temperature
is dependent on the cooling rate of the crystal. Slow cooling
measurements are made after cooling the sample from room temperature
to 50 K over approximately six hours. Alternatively, quenching the
sample is accomplished by plunging it directly into the magnetometer
which is already cooled to 5 K. Using this method the sample is cooled
from room temperature to 10 K within 10 seconds. Slow cooling leads to
a T$_c$ of 32 K versus 26 K for the sample quenched from room
temperature.  However, there is a substantially larger shielding
signal for the quenched sample.  The difference in volume fraction of
the shielding signal indicates that quenching lowers T$_c$ by
inhibiting the phase separation process.  A thorough discussion of the
effects on T$_c$ of quenching $\rm La_2CuO_{4 +\delta}$ is available
elsewhere.\cite{Chou3} The much slower cooling rates available for the
neutron scattering experiments do not allow us to study directly the
effects of quenching on the structure.  It should be noted that the
T$_c$ of sample E is higher than that of the others and the transition
is sharper; we believe that this occurs because the parent $\rm
La_2CuO_4$ crystal is grown using the float zone technique and hence
is purer than the crucible-grown crystals.

\hspace{.20in}  The relative intensities of the two (0,0,8) peaks provide a
second measure of the fraction of the sample in the two phases.  There
may be some error in this calculation because there can be small
variations in the structure factor for the (0,0,8) peak in the two
phases.  A fit of the (0,0,8) peak at low temperatures to two
gaussians indicates that 16\% of sample E is oxygen-rich and 84\% is
oxygen-poor.  This result is comparable to that obtained from the
relative strength of the weak ferromagnetic signals described above,
which gave phase fractions of 20\% and 80\%, respectively.  From the
phase diagram previously determined by Radaelli {\it et al.}
\cite{Radaelli1} for powder $\rm La_2CuO_{4 + \delta}$ samples we
infer a macroscopic oxygen concentration of 4.02 for this sample with
80\% $\delta$=0.012 and 20\% $\delta$=0.055.

\hspace{.20in}  All of the samples we have studied show multiple phases.  Five
samples show coexistence of an oxygen-poor Bmab phase and a staged phase.  Of
these, A1, A2, B and D contain two phases at all temperatures studied.  The
highest measurement temperatures vary:  \ sample A1 (stage 5.5), A2 (stage 4),
and C (stages 2,3,4) have been measured up to 300 K, sample B (stage 4.2) has
been measured up to 320 K, and sample D (stage 5.5) has been studied up to 390
K.  There are two possible reasons why the latter four samples do not form a
uniform phase.  The first is that the phase separation temperature for these
samples is simply higher than the temperatures at which they have been
measured.  The second is that the two phases result from macroscopically
inhomogenous regions in the crystals rather than an equilibrium phase
separation.

\hspace{.20in}  The phase diagrams that have been published by Radaelli {\it et
al.} \cite{Radaelli1} and Hammel {\it et al.} \cite{Hammel} both show
that the phase separation temperature for a sample near the
oxygen-rich side of the miscibility gap is near 280 K.  Because of
this, we believe that the mostly stage 4 samples, A2 and B do not show
true equilibrium phase separation.  Rather, some small region of each
sample is not in equilibrium with the majority of the crystal and
remains oxygen-poor.  It is more likely that samples A1 and D do show
orthodox equilibrium phase separation since they both have substantial
oxygen-poor volume fractions.  The phase separation temperature may be
much higher in these samples.  The phase diagrams determined by
Radaelli {\it et al.}
\cite{Radaelli1} and Hammel {\it et al.} \cite{Hammel}
both indicate phase separation temperatures near 400 K for oxygen
concentrations at the middle of the miscibility gap.  We know that the
stage 6.6 phase represents the structure of $\rm La_2CuO_{4 + \delta}$
at the lowest temperature, oxygen-rich edge of the miscibility gap in
crystal E which was grown by the TSFZ method.  We suspect that the
phase at the oxygen rich side of the miscibility gap for crystals
grown by the top-seeded solution method, as seen in samples A and D,
is stage 5.5 The staging of the phase on the oxygen-rich side of the
miscibility gap thus varies from crystal to crystal, probably because
of the specific defects present in each crystal.

\section{Summary and Conclusions}

\hspace{.20in}  We have used a high temperature electrochemical process to
produce large single crystals of super-oxygenated $\rm La_2CuO_{4 +
\delta}$.  Neutron scattering studies of these crystals reveal that
the oxygen-rich phases in these crystals have a superstructure
different from that observed in undoped or Sr-doped $\rm La_2CuO_4$.
In particular, instead of superlattice peaks at Bmab symmetry
positions, we find pairs of peaks that are displaced by equal amounts
$\pm \Delta$ from the Bmab positions.  We have proposed a model that
attributes these peaks to a staging phenomenon, that is, the excess
oxygen ions are ordered into regularly spaced interstitial planes.
The ordering of the oxygens causes a series of anti-phase domain
boundaries between tilt directions of the $\rm CuO_6$ octahedra.  A
great deal of work remains to be done to understand completely the
$\rm La_2CuO_{4 + \delta}$ staging phase diagram.  The microscopic
structure that we have measured for the oxygen-rich phases of $\rm
La_2CuO_{4 + \delta}$, is quite different from what has been
previously assumed.  It is now necessary to revisit models for hole
driven phase separation in light of our new understanding of the
structures present on both sides of the miscibility gap.

\hspace{.20in} An interesting and important difference between the $\rm
La_2CuO_{4 + \delta}$ and $\rm La_{2-x}Sr_xCuO_4$ superconducting
systems is that in the former, the oxygen dopants give rise to
annealed disorder whereas in the latter, the Sr dopants produce
quenched disorder.  By annealed disorder one means that in $\rm
La_2CuO_{4 + \delta}$ the intercalated oxygen ions reach their
equilibrium configuration at a temperature which, in energy units, is
much less than the characteristic electronic and magnetic energies in
this system.  By contrast in $\rm La_{2-x}Sr_xCuO_4$ the Sr
distribution is determined at the growth temperature and there is no
further equilibration with respect to the $\rm Cu^{2+}$ magnetic or
electronic system at lower temperatures.  This is described as
quenched disorder.  Of course, it seems likely that in neither case is
there equilibration with respect to the superconductivity.  It will be
very valuable to explore any possible differences in these two systems
which reflect the differences in the nature of the dopant disorder.

\hspace{.20in}  Clearly studies of the transport, optical and microscopic
magnetic properties of these novel $\rm La_2CuO_{4 + \delta}$ lamellar
superconductors as a function of stage number would be most
interesting.  Many materials physics challenges remain as well.  For
example, it is essential to characterize quantitatively the
relationship between the excess oxygen concentration and the stage
number.  On the materials preparation side we need to learn how to
prepare larger single crystals with increased homogeneity, large and
uniform staging domains, and sharp superconducting transitions.

\hspace{.20in} The work at MIT was supported by the MRSEC Program of the
National Science Foundation under award number DMR 94-00334 and by the
NSF under award number DMR 93-15715.  Research at Brookhaven National
Laboratory was carried out under contract No.\@ DE-AC-2-76CH00016,
Division of Material Science, U.S. Department of Energy.  Ames
Laboratory is operated for the U.S. Department of Energy by Iowa State
University under contract No. W-7405-Eng-82.  The work at Ames was
supported by the Director for Energy Research, Office of Basic Energy
Sciences.

\newpage

\begin{center}{{\bf Figure Captions}}
\end{center}
\begin{description}

\item [Fig.\@ 1] \ Magnetization as a function of temperature for an applied
field of 0.5 T.  There is a peak in each scan at the N\'{e}el temperature that
is indicative of the weak ferromagnetic transition.  The electrolysis process
was stopped at regular intervals in order to take the magnetization data.  The
data set labeled (1) was taken on the as-grown crystal before oxidation.  Data
set (2) was taken after one week of electrolysis, set (3) after two weeks and
set (4) after four weeks.  The lines are guides to the eye.

\item [Fig.\@ 2] \ Parts a and b are a schematic structure for $\rm La_2CuO_{4
+ \delta}$.  An {\bf a}-axis projection of the real space structure is
shown in (a) for the undoped $\delta$=0 phase and in (b) for the model
stage 4 structure.  The double triangles represent $\rm CuO_6$
octahedra, and the La atoms are left out for clarity.  The dashed
lines indicate octahedra displaced by {\bf a}/2 out of the plane of
the paper with respect to the octahedra drawn with solid lines.
Figure 2c is a reciprocal-space map of the allowed reflections of the
two structures for the [H,0,L]/[0,K,L] scattering zone.  The solid
circles indicate the positions of the fundamental peaks present in
both structures, the open triangles are the Bmab superstructure peaks
for the oxygen-poor phase and the open circles are the stage 4
superlattice peaks for the oxygen-rich phase.

\item [Fig.\@ 3] \ Neutron scattering scans along the reciprocal space L
direction revealing the superlattice peaks detected in various
crystals.  The peak at the (0,1,4) position is a Bmab superlattice
peak and is cut off in the lower panels to emphasize the staging
peaks.  The vertical dashed lines are included for reference at (0,1,4
$\pm \frac{1}{4}$).  The solid curve in the bottom panel is a guide to
the eye.

\item [Fig.\@ 4] \ Scan along L in the vicinity of the (0,3,2) position
for sample B showing superlattice peaks due to both the oxygen-poor,
Bmab phase and the oxygen-rich, stage 4 phase.  This wide scan shows
that the oxygen-rich phase has strong peaks at (0,3,2 $\pm
\frac{1}{4}$), but at best barely detectable peaks at (0,3,2 $\pm
\frac{3}{4}$) with only a very weak signal at (0,3,1$\frac{1}{4}$) and
no intensity above background at (0,3,2$\frac{3}{4}$).

\item [Fig.\@ 5] \ Lattice parameters as a function of temperature for several
samples of $\rm La_2CuO_{4 + \delta}$.  The top panel shows the {\bf c} lattice
constant and the lower panel shows the in-plane lattice constants.  The open
symbols are for the minority oxygen-poor phases and the solid symbols are for
the oxygen-rich phase and cannot be independently measured.

\item [Fig.\@ 6] \ The temperature dependence of the out-of-plane half-width at
half-maximum (HWHM), in-plane HWHM, and the integrated peak intensity
for the (0,3,1$\frac{3}{4}$) peak for two different samples with stage
4 phases.  The HWHM and intensities were determined by fits to the
peaks of a 3D Lorentzian squared profile convolved with the
instrumental resolution.

\item [Fig.\@ 7] \ Magnetization for sample E as a function of
temperature.  The top panel shows a high-field scan revealing the weak
ferromagnetic transition of the oxygen-poor phase and the lower panel
show zero field cooled measurements of the shielding signal of the
oxygen-rich superconducting phase.

\item [Fig.\@ 8] \ Neutron scattering scans at room temperature (open circles)
and 10K (filled circles) showing the effects of phase separation and
stage ordering.  The top panel shows the emergence of the (0,1,4 $\pm$
0.15) staging peaks at low temperatures.  The bottom panel shows the
(0,0,8) peak revealing a single {\bf c}-axis at room temperature which
is split to reveal two phases with different {\bf c}-axes at low
temperatures.

\end{description}

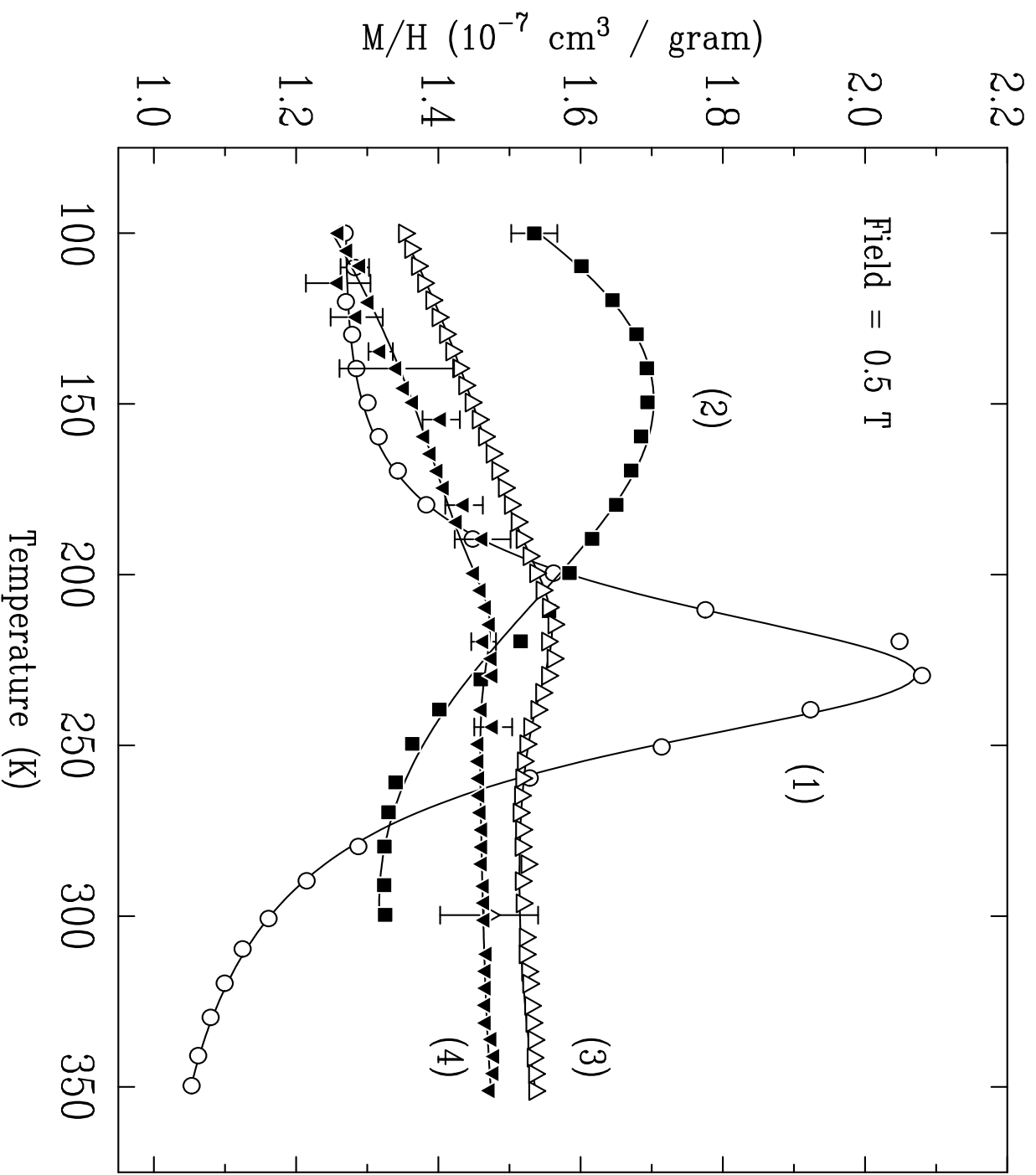
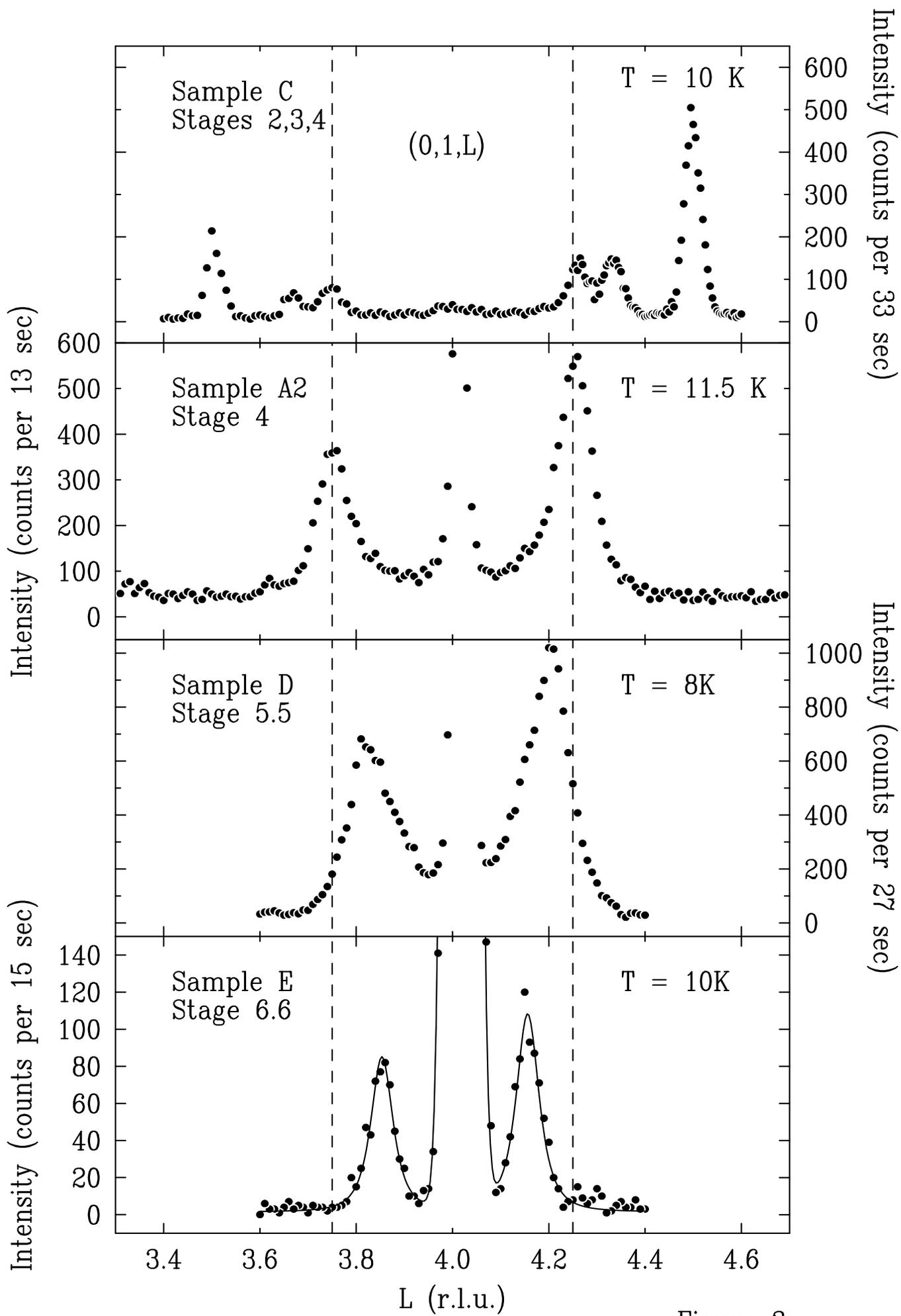
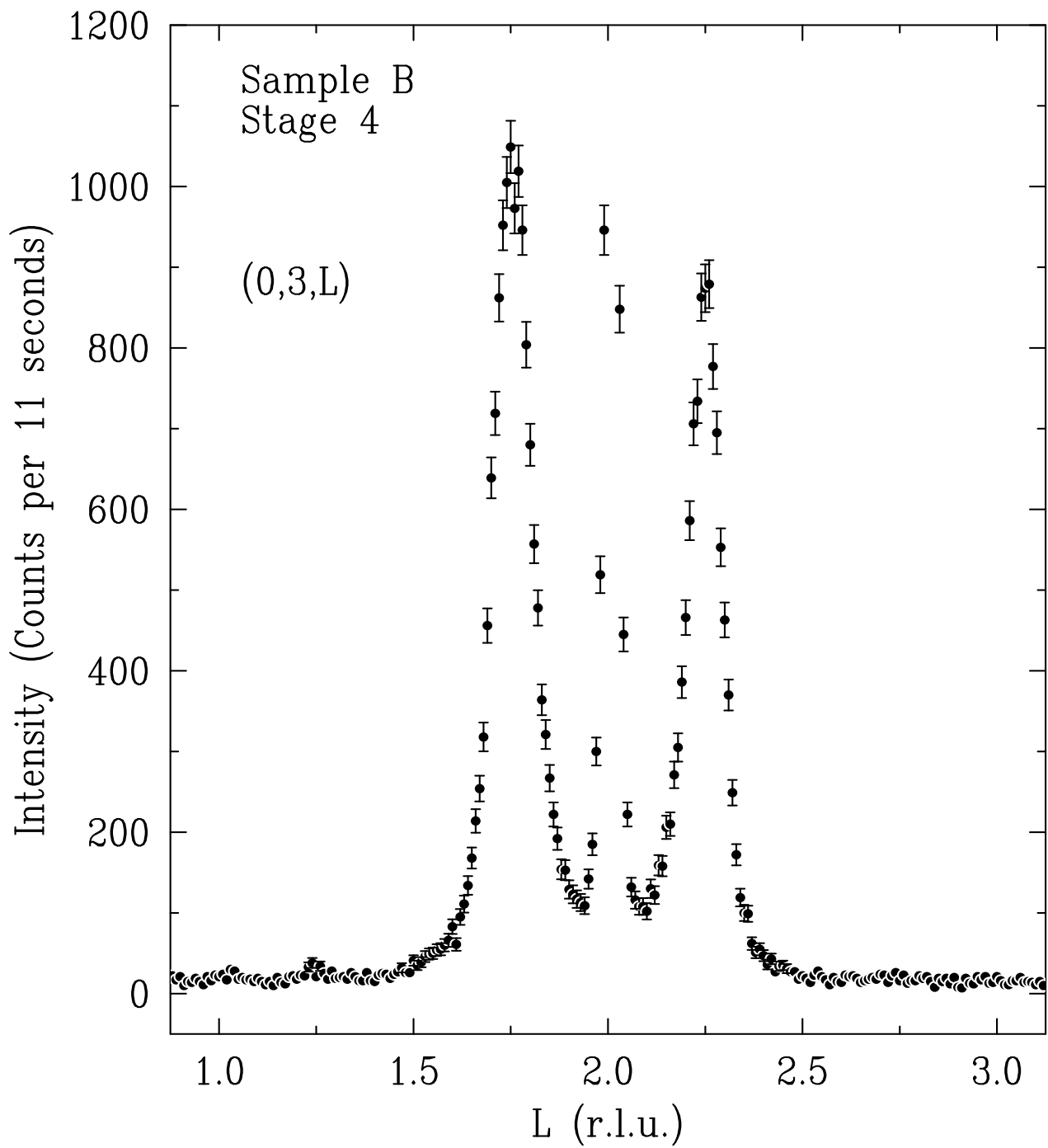
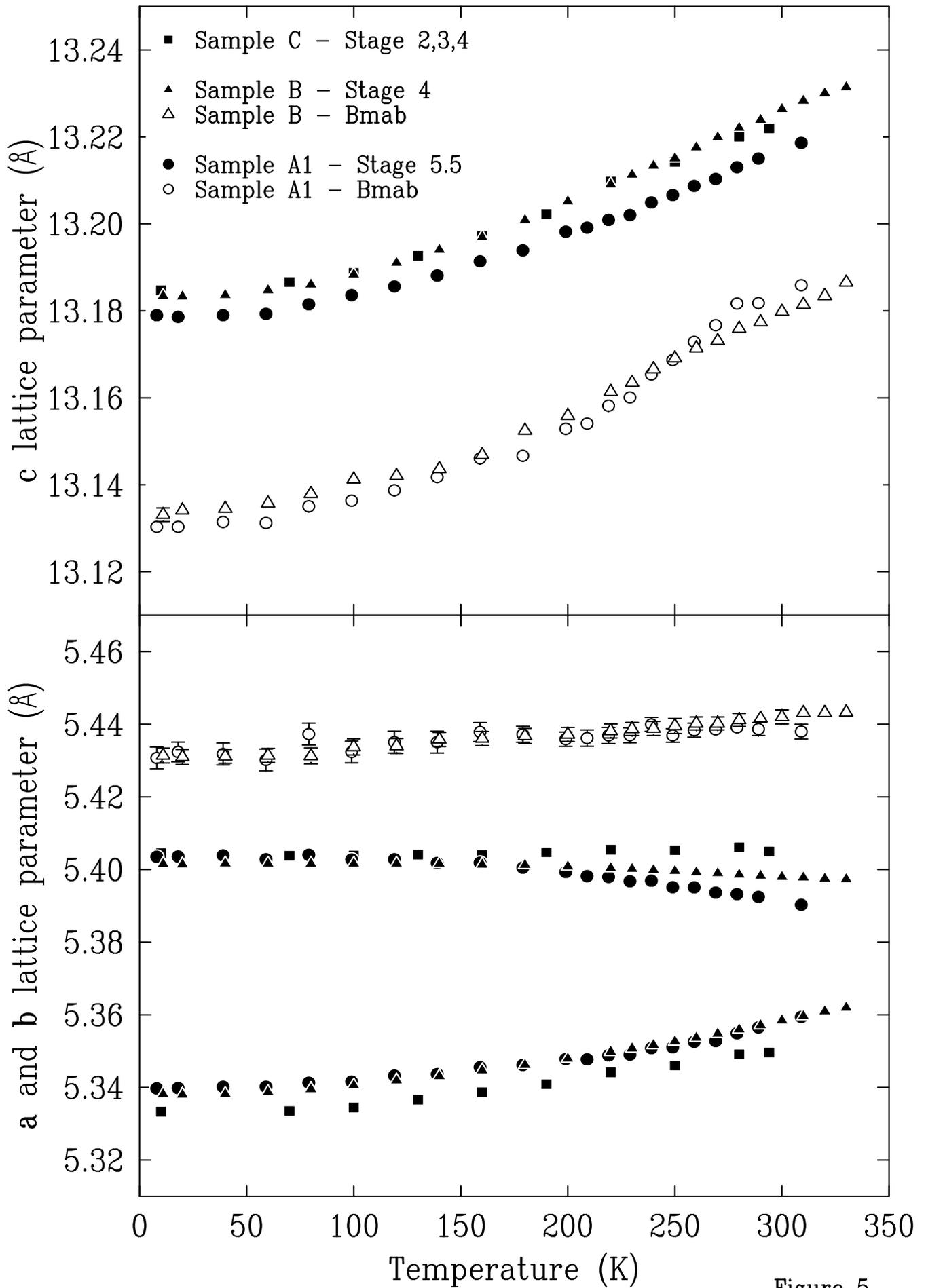
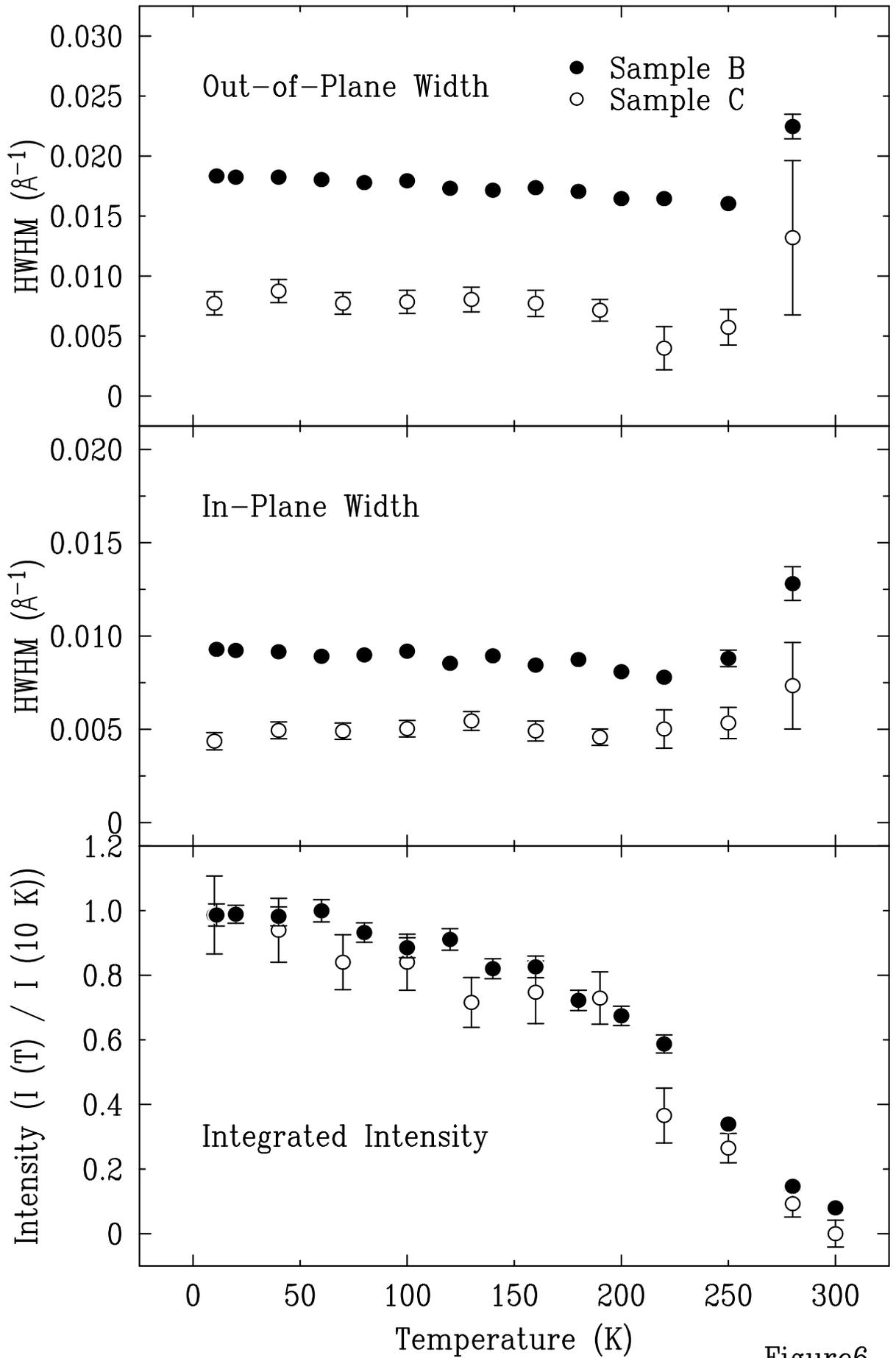
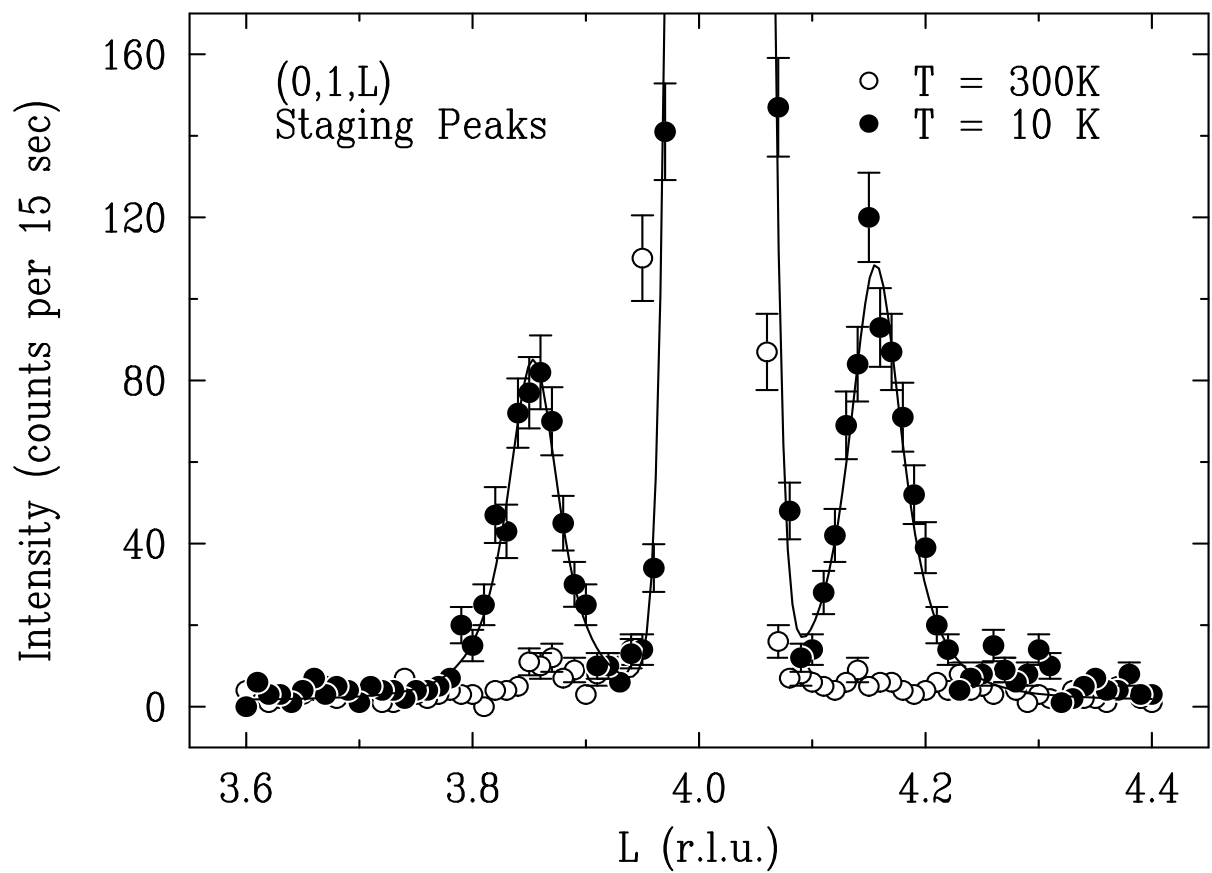
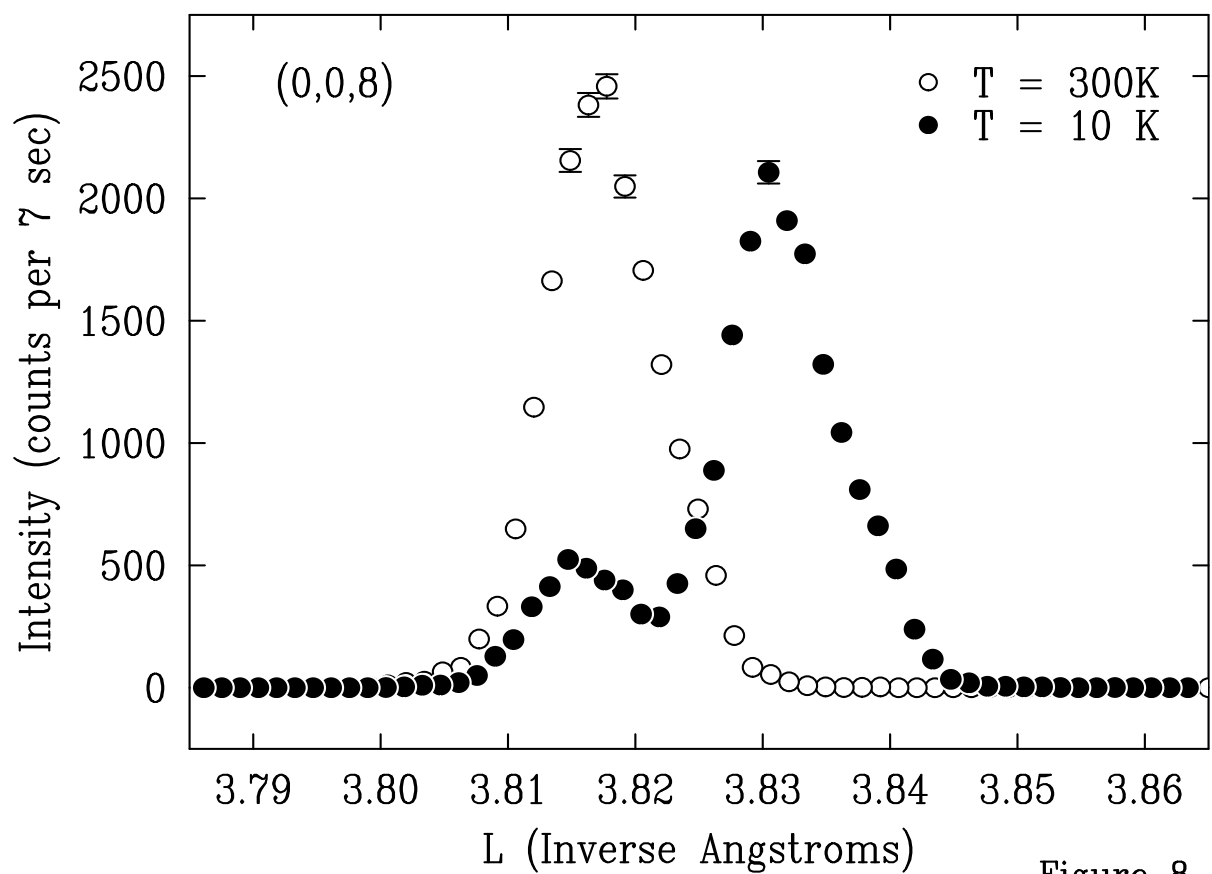
\end{document}